# On-chip radially and azimuthally polarized vortex beam generation utilizing shallow-etched gratings

ZENGKAI SHAO,[1] JIANGBO ZHU,[2] YANFENG ZHANG,[1,*] YUJIE CHEN,[1] AND SIYUAN YU[1,2,*]

[1]State Key Laboratory of Optoelectronic Materials and Technologies, School of Electronics and Information Technology,
Sun Yat-sen University, Guangzhou 510275, China
[2]Photonics Group, Merchant Venturers School of Engineering, University of Bristol, Bristol BS8 1UB, UK
*Corresponding author: zhangyf33@mail.sysu.edu.cn; s.yu@bristol.ac.uk

**Cylindrical vector vortex (CVV) beams, complex light fields that exhibit a vector nature and carry quantized orbital angular momentum states, have been widely investigated due to their rich applications. Current technologies to generate CVV beams using individual polarization and spatial phase manipulations suffer from bulky size and low configurability. In this work, we propose and experimentally demonstrate an approach to generate CVV beams with an integrated device based on silicon nitride microring resonator and embedded top-gratings. The device is capable of generating broad-band radially and azimuthally polarized vortex beams by simply switching the polarization and wavelength of the excited whispering-gallery-modes. In addition, we develop a method to fabricate the devices of shallow-etched top-gratings with only one-step etching. This novel method provides new capabilities to develop on-chip integrated devices with great ease and flexibility. © 2017 Optical Society of America**

Cylindrical vector (CV) beams are vector solutions of Maxwell's equations with an inhomogeneous but cylindrically symmetric polarization distribution in the transverse plane. Radially polarized (RP) and azimuthally polarized (AP) beams are distinctive CV beams that have a wealth of applications in imaging [1, 2], quantum memories [3] and optical manipulation [4-6]. Meanwhile, optical vortex beams with spiral wave fronts have also attracted much interest as they carry intrinsic orbital angular momentum (OAM) of light [7]. Cylindrical vector vortex (CVV) beams combining the characteristics of spatial polarization variation and helical phase fronts enable a variety of new features such as photonic spin Hall effect [8] and sharper focusing [9], and can find applications in laser micro-processing [10]. Many approaches to generating vector vortex beams use bulk optical elements such as q-plates, polarizers, wave plates and spiral phase plates. Recently, more compact approaches such as meta-surface based [11] and surface-grating based [12-16] devices have been demonstrated to generate CVV beams. A meta-surface device in principle only generates a CVV beam with fixed OAM order and polarization profile. A silicon micro-ring based vortex beam emitter is recently developed, with sub-wavelength scatterers embedded along the inner-sidewall of the ring resonator in a 2nd-order grating manner and coupling the whispering-gallery modes (WGMs) in the resonator to free-space propagating vortex beams [12]. High compactness, reliability, and tunability in the emitted topological charge can be achieved [12, 15]. However, these emitters typically generate vortex beams in near-azimuthal polarization, due to the strong tangential field at the high contrast waveguide boundary between silicon and cladding dielectrics ($SiO_2$ or air). In [16] a shallow-ridge silicon micro-ring is theoretically analyzed to generate RP vortex beams by exploiting shallow-etched holes at the top of the waveguide. Meanwhile, CVV emitters of high configurability in both topological charge and polarization state are still highly desirable for many practical applications.

In a recent work, we have demonstrated the interaction between transverse spin angular momentum in evanescent waves and the intrinsic OAM in optical vortex beams by tailoring the silicon nitride ($SiN_x$) micro-ring waveguide [17]. Vortex beams with a large range of polarization states are generated with this device by engineering the transverse spin in evanescent waves. However, generating high purity RP or AP vortex beams is still challenging because the azimuthal component $E_\varphi$ and radial component $E_r$ are generally comparable at the sidewalls of $SiN_x$ waveguides for TE mode. In this work, we experimentally demonstrate another approach to generating CVV beams using a single $SiN_x$ micro-ring resonator with shallow-etched periodic angular gratings placed on top of the ring waveguide. As opposed to [16], this device can be configured to generate both RP and AP cylindrical vector beams of well-defined OAM states. In addition, the angular top-gratings are shallow-etched in a one-step process, which simplifies the complexity of fabrication process. The proposed approach provides an extremely simple and flexible on-chip solution to generating CVV beams.

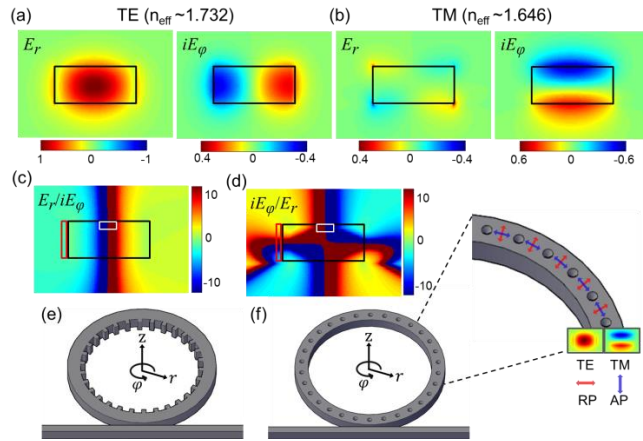

Fig. 1. Simulated cross-sectional field distributions of the radial component $E_r$ and azimuthal component $E_\varphi$ for (a) TE and (b) TM mode in a $SiN_x$ waveguide at 1550 nm. The corresponding distribution of the component ratio (c) $E_r/iE_\varphi$ for TE mode and (d) $iE_\varphi/E_r$ for TM mode. Schematic of the designed vortex beam device in our previous work (e) and in this work (f).

WGMs in a ring cavity carry high-order OAM states [18]. By introducing angular gratings or download waveguides array [19], the confined WGMs in the ring cavity can be coupled into radiative cylindrical modes. In addition, the state of polarization (SoP) of the radiated CVV beams is determined by the SoP of the local optical field being scattered at each grating element [12]. Using a finite difference eigenmode (FDE) solver, we simulated the cross-sectional electric field distribution of fundamental TE mode and TM mode in a $SiN_x$ waveguide, as shown in Figs. 1(a) and 1(b). The designed $SiN_x$ micro-ring waveguide has a cross-section of 1.2 μm in width and 580 nm in height in order to achieve strong confinement for both TE mode and TM mode. The waveguide is fabricated on a 5-μm-thick thermal-oxide silicon substrate and surrounded by air. Silicon nitride has a moderate refractive index of ~2.0, affording high fabrication tolerance and low phase sensitivity. The calculated effective refractive indices are $n_{eff}$ = 1.723 for fundamental TE mode and $n_{eff}$ = 1.646 for fundamental TM mode at $\lambda$ = 1550 nm. For the TE mode, apart from $E_r$, a strong azimuthal component $E_\varphi$ at the lateral core-cladding interface (sidewalls) can also be observed in $\pm\pi/2$ phase difference to $E_r$, while $E_\varphi$ is negligible at the center of the waveguide. For TM mode, strong $E_\varphi$ exists at the upper and lower core-cladding interfaces in $\pm\pi/2$ phase difference to $E_r$, while $E_r$ is negligible at the center of the waveguide. Figs. 1(c) and 1(d) show the calculated $E_r/iE_\varphi$ for TE mode and $iE_\varphi/E_r$ for TM mode, respectively. In our previous work [15], the angular grating elements are located at the inner sidewall of the waveguide, shown in Fig. 1(e), where the amplitude of $E_r$ and $iE_\varphi$ are comparable ($E_r/iE_\varphi$ ~ 1.25) for TE mode [marked in Fig. 1(c) by the red box]. It is very difficult to eliminate one of electric components by tailoring waveguide dimensions to generate pure RP or AP vortex beams.

With the local field of the TE and TM modes at the upper core-cladding boundary of the waveguide dominated by $E_r$ and $E_\varphi$, respectively, placing the gratings at the top [e.g., within the region marked by the white box in Figs. 1(c) and 1(d)] allows for the generation of AP and RP vortex beams by simply switching the mode (i.e., TE or TM) in the ring. In addition, the topological charge of radiated vortex beam $l$ satisfies the selection rule of $l = p - q$ (where $p$ is the WGMs order in the ring, $q$ is the number of grating elements around the resonator), which can be tuned by simply changing the input wavelength so that it aligns with different ring cavity resonances. In this work, the device for demonstration comprises of a $SiN_x$ ring resonator with a radius of 60 μm. The gap between the access waveguide and the ring is fixed as 210 nm. Shallow-etched holes are used as periodic angular scatterers on the top of the waveguide, shown in Fig. 1(f). With these scatterers arranged in a second-order grating fashion, the first-order diffracted light from the WGMs collectively produce a vortex beam carrying optical OAM and propagating perpendicular to the resonator plane, with a SoP of RP or AP depending on the mode in the ring waveguide being TE or TM.

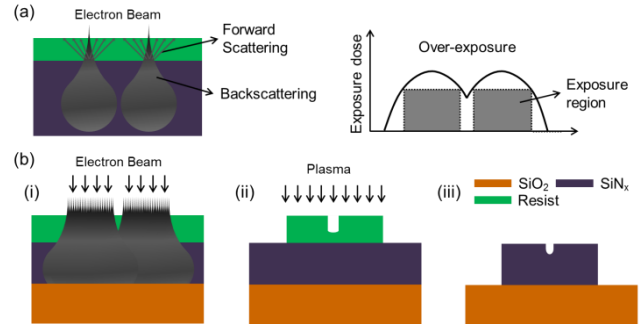

Fig. 2. (a) Left: Schematic of the electrons distribution during the E-beam lithography resulting from forward scattering and backscattering predominantly. Right: Curve of the exposure dose with over-exposure of the gap between waveguides. (b) Fabrication process of the proposed structure. (i) Exposure of the patterns by E-beam lithography. (ii) RIE process after development of resist. (iii) Lift-off the resist.

In general, the proposed device with shallow-holes as a three-dimensional microstructure can be fabricated by many technologies, such as gray-scale lithography [20, 21], photoresist reflow processing [22], and multi-step etching [23]. However, those methods increase the complexity of the fabrication process and have special requirements for the photoresist. Here, we develop a simple method to etch through the 580-nm thick $SiN_x$ waveguide with the embedded shallow-holes in a one-step process utilizing a combination of the over-exposure of negative resist resulting from forward scattering and backscattering of electrons during the E-beam lithography (EBL) [24] [shown in Fig. 2(a)] and the lag effect [25] during the reactive-ion etching (RIE) process. Figure 2(b) shows details of the fabrication process. In our experiment, the 580-nm $SiN_x$ layer is deposited on a thermally oxidized <100> silicon wafer with an oxide thickness of 5 μm using an inductively-coupled plasma chemical vapour deposition (ICP-CVD) process [26]. A 480-nm-thick negative resist (diluted AZ nLOF 2035 from Clariant Corporation) is used to pattern structures with the EBL system at 100 kV.

To achieve precisely controlled shallow-etched holes, a series of slot waveguides with different gap widths (from 80 nm to 300 nm in 20-nm steps) are examined under an optimized exposure dose of 170 μc/cm$^2$. Figures 3(a) and 3(b) show the over-all view and zoom-in cross-section of photoresist profiles after development, respectively. It can be seen in Fig. 3(c) that under such conditions, gaps of 180 nm and above can be fully developed and etched through by the subsequent RIE process. Patterns of 120 nm- and 140 nm-gap can be distinguished with obvious residual photoresist even after RIE etching, therefore leaving the $SiN_x$ unetched. When the gap is 160 nm, appropriate over-exposure is achieved resulting in a shallow-etched pattern into the $SiN_x$ after RIE [marked by a red box in Fig. 3(b) and (c)]. A smooth and narrower profile at the bottom of the etched slot is observed due to the lag-effect during the RIE process, which helps to reduce the perturbation on the field distribution in the waveguide and reduce the propagation loss of both TE and TM modes. Figure 3(d) shows the zoom-in view of a CVV emitter fabricated by this method. The number of periodic shallow-holes is fixed at $q$ = 420 with the calculated zero topology charge ($l$ = 0) at $\lambda$ = 1550 nm for TE mode and $\lambda$ = 1495 nm for TM mode.

The devices for characterization and demonstration here are all with shallow-etched holes of 160 nm wide. However, placing holes of this size at the top center results in a decaying emitted intensity along the ring [as shown in Fig. 3(e) with d = 0], indicating a grating coupling strength that is too strong. Such intensity inhomogeneity would degrade the mode purity. To achieve a decent profile as well as a high emission efficiency, devices with holes of various offsets (d) to the center of waveguide are fabricated and compared, as shown in Fig. 3(e). A uniform angular intensity distribution is obtained at d = 60 nm, which is employed for devices in the following experiments.

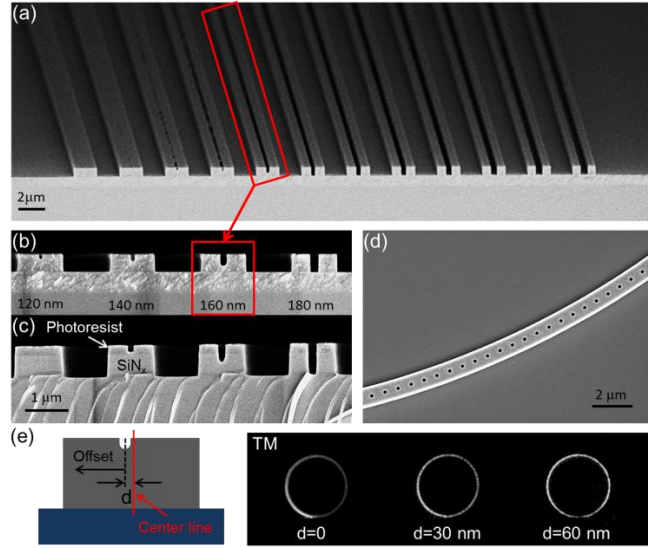

Fig. 3. (a) SEM image of the photoresist patterns after the E-beam lithography and development. Zoom-in cross-section of patterns (b) before and (c) after etching through a 580-nm $SiN_x$ layer by the RIE process. (d) Zoom-in view of the fabricated device. (e) Schematic of the local offset of the shallow-etched holes on the top of the waveguide and the measured near-field intensity when d=0, 30 nm, 60 nm, respectively.

Figure 4(a) shows the emission spectral response of the vortex beam emitter obtained by scanning the wavelength of the input laser (1530-1580 nm) when exciting TE or TM WGMs in the resonator, respectively. A polarization controller (PC) is used to set the input polarization to TE or TM mode. The mode splitting near 1553.4 nm for TE mode and near 1510.4 nm (not shown in the figure) for TM is observed, which is the signature of $l$ = 0, due to the Bragg back-reflection of the grating. These wavelengths are consistent with the simulation results, demonstrating the high fabrication tolerance of $SiN_x$ devices. The average emission efficiency is calculated to be 7.0% for TE mode and 9.2% for TM mode. The electric field magnitude of radial component $E_r$ for TE mode shown in Fig. 1(b) is mainly concentrated in the center of the waveguide cross-section, resulting in relative low intensity near the top of the waveguide, hence low emission efficiency. Note that emission power is decreased at the resonance wavelength near $l$ = 0 for TE mode, which is the result of strong scattering and coupling between adjacent modes. The emission efficiency can be increased by optimization of the grating dimensions, the access waveguide coupling ratio (so that critical coupling is achieved), or by introducing a reflective mirror at the bottom of the buried oxide layer.

The first pictures on the left of Figs. 4(b) and 4(c) show the near-field intensity profiles of the generated vortex beam for TE and TM mode, respectively. The polarization states of the vortex beams are examined using a linear polarizer. The measured intensity distributions after passing through an analyzing polarizer in horizontal, 45°, vertical, and −45° orientations are shown in Figs. 4(b) and 4(c). The intensity distributions of the two symmetric lobes for TE (TM) mode are parallel (perpendicular) to the trans-axis of the polarizer, confirming that the vortex beam is radially (azimuthally) polarized.

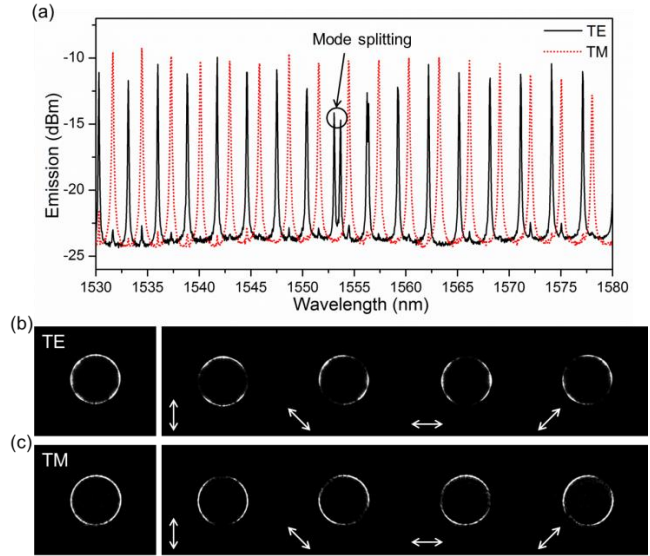

Fig. 4. (a) Emission spectrum measured by scanning the input light wavelength with switching the polarization of excited WGMs. Measured near-field intensity patterns after passing through a polarizer with different polarization angles for (b) TE mode and (c) TM mode. The polarization angles are denoted by double-headed arrows in white.

To further accurately confirm the polarized component of the radiated CVV beams, a radial polarization converter (RPC) is utilized to measure the radial and azimuthal electric components when the input light is TE or TM mode, respectively [27]. Figure 5(a) shows the measurement setup consisting of a RPC and a linear polarizer (LP). Passing through the RPC element, the two orthogonal components (RP and AP) are rotated to linear polarization state by twisted nematic liquid-crystal, which is parallel or perpendicular to the alignment layer (white dot line). Followed by a rotatable polarizer, the RP and AP components of the radiated CVVs beam are filtered in 0° and 90° orientations, and then imaged by an infra-red camera. Figure 5(b) and 5(c) show the measured intensity distributions for TE and TM mode in 0° and 90° orientations, respectively. It's evident that the emitted beam by launching TE (TM) mode is predominantly transformed into horizontally (vertically) linear-polarized, indicating the generation of the RP and AP vortex beams.

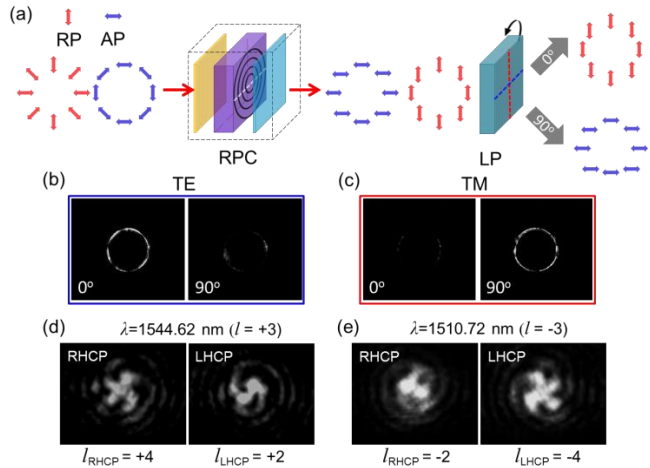

Fig. 5. (a) The schematic diagram of the measurement setup. The near-field intensity distributions of the vortex beam through the RPC element and linear polarizer at the position of 0° and 90° for (b) TE and (c) TM mode, respectively. Interference patterns are measured for (d) TE mode at $\lambda$= 1544.62 nm with $l$ = +3 and (e) TM mode at $\lambda$= 1510.72 nm with $l$ = -3.

In addition, the RPC element is replaced by the circular polarization filter consisting of a quarter-wave plate and a linear polarizer to reveal the spiral wavefront and verify the topological charge of optical vortex. It has been demonstrated that the AP and RP vortex beams can be described as the superposition of two orthogonal scalar vortices which consist of a right-hand circularly polarized (RHCP) beam with topological charge of ($l$ + 1) and a left-hand circularly polarized (LHCP) beam with ($l$ – 1) [28]. The value and sign of $l$ would be observed by interfering the generated beam with a copropagating LHCP or RHCP Gaussian beam as reference [12]. For the coupled TE mode at $\lambda$ = 1544.62 nm ($l$ = +3) and TM mode at $\lambda$ = 1510.72 nm ($l$ = -3), the measured interference patterns with RHCP and LHCP reference beams are shown in Figs. 5(b) and 5(c), respectively. The number of spiral arms agrees well with the prediction for both TE and TM modes.

In summary, we have demonstrated an integrated device that emits cylindrical vortex beams based on a SiN$_x$ micro-ring with shallow-etched angular gratings on the top of the ring waveguide. The integrated device, which exploits the large difference of the radial and azimuthal electric distribution between excited TE and TM modes in the waveguide, is capable of generating radially and azimuthally polarized vortex beams by simply switching the polarization of the input light. Our device provides an extremely simple and feasible approach for an integrated cylindrical vortex beam switch, opening a new path for practical applications of the structured beams in nanoparticle trapping, optical communications, and quantum optics.

**Funding.** National Basic Research Program of China (973 Program) (2014CB340000); National Key Research and Development Program of China (2016YFB0402503); National Natural Science Foundations of China (61490715, 61323001, 11774437, 11690031, 51403244); Natural Science Foundation of Guangdong Province (2014A030313104); Science and Technology Program of Guangzhou (201707020017); Fundamental Research Funds for the Central Universities of China (Sun Yat-sen University: 17lgzd06, 16lgjc16).